\documentclass[aps,jcp,reprint,superscriptaddress,showkeys]{revtex4-1}
\usepackage{graphicx}
\usepackage{amsmath}
\usepackage{amssymb}
\usepackage{longtable}
\usepackage{dcolumn}
\usepackage{natbib}
\begin{document}

\title{Electrostatic and induction effects in the solubility of water in alkanes}
\author{D. Asthagiri}\thanks{Email: dna6@rice.edu}
\affiliation{Chemical and Biomolecular Engineering, Rice University, Houston, TX} 
\author{A. Valiya Parambathu}
\affiliation{Chemical and Biomolecular Engineering, Rice University, Houston, TX} 
\author{Deepti Ballal}
\affiliation{Materials Science and Engineering, Ames Laboratory, Ames, IA}
\author{Walter G.~Chapman}\thanks{Email: wgchap@rice.edu}
\affiliation{Chemical and Biomolecular Engineering, Rice University, Houston, TX}

\date{\today}
\begin{abstract}
Experiments show that at 298~K  and 1 atm pressure the transfer free energy, $\mu^{\rm ex}$,  of water from its vapor to liquid normal alkanes $C_nH_{2n+2}$ ($n=5\ldots12$) 
is negative. Earlier it was found that with the united-atom TraPPE model for alkanes and the SPC/E model for water, one had to artificially enhance the 
attractive alkane-water cross interaction to capture this behavior. Here we revisit the calculation of $\mu^{\rm ex}$ using the polarizable AMOEBA  and the non-polarizable Charmm General (CGenFF) forcefields. We test both the AMOEBA03 and AMOEBA14 water models; the former has been validated with the AMOEBA alkane model
while the latter is a revision of AMOEBA03 to better describe liquid water.  We calculate $\mu^{\rm ex}$ using the test particle method. With CGenFF, $\mu^{\rm ex}$ is positive and the error relative to experiments is about 1.5~$k_{\rm B}T$.  With AMOEBA, $\mu^{\rm ex}$ is negative and deviations relative to experiments are between 0.25~$k_{\rm B}T$ (AMOEBA14) and 0.5~$k_{\rm B}T$ (AMOEBA03).  Quantum chemical calculations in a continuum solvent suggest that zero point effects may account for some of the deviation.  Forcefield limitations notwithstanding, electrostatic and induction effects, commonly ignored in considerations of water-alkane interactions, appear to be decisive in the solubility of water in alkanes. 
\end{abstract}


\maketitle

\section{Introduction}

Alkane solubility in water is of fundamental importance in understanding various self-assembly processes and of the nature of water itself \cite{stillinger:sc80,lrp:jpcb98,chandler:nature05}.  This process has been studied extensively over the last few decades both experimentally and in computer simulations. The converse process, the solubility of water in alkanes, has received considerably less attention.  Nevertheless, understanding the solubility of water in alkanes is also of fundamental scientific and technological interest. Technologically, one encounters
the problem of characterizing water solubility in alkanes in all aspects related to water in oil and oil-related products. Scientifically, this problem challenges our appreciation of the intermolecular forces between alkane, a nonpolar substance, and water, a prototypical polar compound. The present paper is motivated by these scientific and technological considerations. 

In discussions of alkane solubility in water, the emphasis is typically on the thermodynamics of creating a cavity in the liquid \cite{lrp:jpcb98,chandler:nature05}, with the different physical ingredients of the attractive interaction between the alkane and the water lumped into a single alkane-water dispersion interaction. Usually the dispersion interaction is
modeled using the Lennard-Jones potential with the Lorentz-Berthelot mixing rule. In many instances it has often been found necessary to enhance the alkane-water attractive interaction over that based on using the Lorentz-Berthelot mixing rule (for example, see Refs.\ \cite{Docherty:jcp06,Ashbaugh:hh}), an approach that is rationalized as a way to account for the physics of polarizability, if polarizability is not explicitly considered. The idea of non-negligible electrostatic interaction between an alkane and water  is almost never considered, but quantum chemical calculations \cite{Szal:jcp05}  do suggest that the bonding between methane, a prototypical hydrophobe, and water, in specific configurations can even be characterized as ``weak" hydrogen bonding. 

Recently, Ballal et al.\ \cite{ballal:jcp14,ballal:jcp16} investigated the solubility of water in alkanes using molecular simulations. In their approach they used the united atom
TraPPE \cite{trappe} model of alkanes and the SPC/E \cite{spce} model for water.  They found that with the Lorentz-Bertholet mixing rule for the alkane-water cross-interaction, the predicted transfer free energy of water to alkanes  $C_nH_{2n+2}$ ($n=5\ldots12$) was much too positive. To recover the experimentally observed fraction of water in alkanes they had to increase the water-alkane attractive interaction. Ballal et al.\ \cite{ballal:jcp14} considered several physical explanations for the enhanced effective attraction, but a compelling explanation for the observed solubility of water in alkanes is still wanting.  As McDaniel and Yethiraj \cite{yethiraj:jcp15} subsequently noted in an insightful comment, the work by Ballal et al.\cite{ballal:jcp14} ``uncovered a wonderful example of where pure-component parameters fail for mixtures" and ``a more accurate model must better describe the electrostatic, induction, and dispersion interactions, all of which are potentially important." Recent studies on water/CO$_2$ and water/n-alkane mixtures further
support this idea \cite{thanos:jpcb17}.

Here we revisit the problem of water solubility in alkanes using the polarizable AMOEBA \cite{amoeba03,amoeba14,amoebaOrg} all-atom forcefield. An important feature of the AMOEBA model is extensive reliance on \textit{ab initio} calculations on small clusters during the parameter development.  Further the electrostatic interactions are 
based on distributed multipoles, which are well known to encode a high level of chemical realism \cite{dma}.  Having a better description of the electric field around the molecule 
is also expected to better describe the physics of induction. On the basis of such a model, we find that both electrostatics and induction effects play a decisive role in describing the solubility of water in alkanes. 

\section{Theory}
\subsection{Analysis of experimental data}\label{sc:expt}
The experimental results \cite{polak,roddy,schatzberg} of the solubility of water in alkanes are reported as a mole fraction ($x_w$) or p.p.m.\ by weight. The alkanes are at 298.15~K,  and while it is not explicitly stated, one can infer from the experimental procedure that the pressure is 1~atm.\ For the temperature and pressure conditions, the pure alkane density is well established and 
we obtain those values from the NIST database \cite{nist}. 

Let the number density of neat alkane be $\rho_a$. Since the mole fraction of water in alkane is of the order of 10$^{-4}$, to an excellent approximation 
the number density of water in the alkane phase is $\rho_w^{(a)} = x_w \cdot \rho_a$. (In this limit, the p.p.m.\ measure is also readily converted to $x_w$.) 
At 298.15~K, using the pressure-volume data from the NIST database \cite{nist},  an estimate of the Poynting pressure correction shows that 
the chemical potential of water at 1~atm.\ pressure and 298.15~K is only about 0.001~kcal/mol higher than that at saturation. Hence for simplicity we will assume  
saturation conditions. Since the alkane solubility in water is also low, we assume that the chemical potential of water in the water-rich phase is the same as that of neat liquid water. Equating the chemical potential of water in the alkane-rich and liquid-water phases we have 
\begin{eqnarray}
\beta[\mu^{\rm ex}_{w|a} - \mu^{\rm ex}_{w|l}] & = & \ln \frac{\rho_w^{(l)}}{\rho_w^{(a)}} = \ln \frac{\rho_w^{(l)}}{x_w \cdot \rho_a} \, .
\label{eq:w2alk}
\end{eqnarray}
$\beta = 1/k_{\rm B}T$, where $k_{\rm B}$ is Boltzmann's constant, is the reciprocal temperature in energy units. $\mu^{\rm ex}_{w|l}$ is the excess chemical potential of water in liquid water and $\mu^{\rm ex}_{w|a}$ is the corresponding quantity in the alkane. The excess chemical potential collects all contributions due to intermolecular interactions and is the quantity of principal interest here.  Please note that mass dependent effects are implicitly considered when we use experimental densities on the right hand side 
of Eq.~\ref{eq:w2alk}. Within classical statistical mechanics, we assume mass dependent contributions cancel in composing the difference on the left hand side of Eq.~\ref{eq:w2alk}. (We return to this point when we consider possible corrections due to quantization of nuclear motion.)

Analogous to Eq.~\ref{eq:w2alk}, 
\begin{eqnarray}
\beta[\mu^{\rm ex}_{w|l} - \mu^{\rm ex}_{w|v}] & = & \ln \frac{\rho_w^{(v)}}{\rho_w^{(l)}} 
\label{eq:w2v}
\end{eqnarray}
gives the free energy to transfer water from its saturated vapor to the saturated liquid.  Thus, from Eqs.~\ref{eq:w2alk} and~\ref{eq:w2v}, the free energy to transfer water from its vapor to the alkane is 
\begin{eqnarray}
\beta[\mu^{\rm ex}_{w|a} - \mu^{\rm ex}_{w|v}] =  \ln \frac{\rho_w^{(v)}}{x_w \cdot \rho_a} \, .
\label{eq:v2a}
\end{eqnarray}
All the quantities on the right hand side of Eq.~\ref{eq:v2a} are known experimentally, and thus $\beta[\mu^{\rm ex}_{w|a} - \mu^{\rm ex}_{w|v}]$, which is the target of
molecular simulations, is also well-characterized.   

For the temperature and pressure under consideration, the water vapor can be treated as an ideal gas, i.e.\ $\beta \mu^{\rm ex}_{w|v} = 0$. To confirm this please note that 
to the lowest order in density $\beta \mu^{\rm ex}_{w|v} = 2B_2\rho_w^{(v)} \approx -3\times 10^{-3}$, where $B_2 = -1160$~cm$^3$/mol \cite{harvey:waterb2} 
is the experimentally determined second virial coefficient of water at 298.15~K. We explicitly note the second virial correction anticipating our discussion below on
the possible role of dimerization of water in the alkane. 

\subsection{Calculation of $\mu^{\rm ex}_{w|a}$}
From the potential distribution theorem \cite{widom:jpc82,lrp:apc02,lrp:book,lrp:cpms} we have 
\begin{eqnarray}
\beta\mu^{\rm ex}_{w|a} = -\ln \langle\langle e^{-\beta\Delta U}\rangle\rangle_0 \, .
\label{eq:pdt}
\end{eqnarray}
$\Delta U  = U_{n+1} - U_{n} - U_{1}$ is the binding energy of the test particle indicated by `1'. $U_{n}$ is the potential energy of the neat solvent, $U_{n+1}$ is the
potential energy of the neat solvent plus the added solute (water), and $U_{1}$ is the potential energy of the test particle. The  outer-average $\langle \ldots\rangle_0$ indicates
that the solvent and solute are thermally uncoupled and the inner-average $\langle\ldots\rangle$ indicates sampling over the conformations of the solute uncoupled from the solvent. 

For the solvation of the rigid TIP4P/2005 \cite{tip4p} water, we have only  a single conformation to consider. In the case of water modeled by the AMOEBA
forcefield, sampling over solute conformations is, in principle, required. However, estimation of bond and angle fluctuations from the known force constants suggests that
to an excellent approximation using the equilibrium conformation of the isolated AMOEBA water is adequate. (Importantly, the binding energy calculations are
expected to be insensitive to the tiny changes in the solute conformation.)  Finally we note that a direct application of Eq.~\ref{eq:pdt} always requires care. 
Here the free energies of interest are all within about $\pm 2$~k$_{\rm B}T$ (see below), and we expect a direct application of Eq.~\ref{eq:pdt} to ensure statistically converged results. Our results below support this expectation.

\section{Methods}
\subsection{Simulations}
We study the solvation of TIP4P/2005 water in normal alkanes $C_nH_{2n+2}$ ($n=5\ldots12$) modeled using CGenFF, the CHARMM General Forcefield \cite{cgenff}, and the solvation of AMOEBA water \cite{amoeba03,amoeba14} in alkanes modeled by the AMOEBA forcefield \cite{amoebaOrg}. For the temperature of 298.15~K and 1 atm pressure, we obtained the density of the alkanes from the NIST database (Table~\ref{tb:density}). We built reference structures for $C_nH_{2n+2}$ ($n=5\ldots12$) using Avogadro \cite{avogadro} and then using Packmol \cite{packmol} packed $N$ (Table \ref{tb:density}) of the alkane molecules in a cubic box to achieve the prescribed density. This system defined the starting configuration. 
\begin{table}[h!]
\caption{Simulation cell density and number of particles. The density $\rho_a$ is obtained from the NIST database \cite{nist}. $N$ is the number of molecules in the system.}
\label{tb:density}
\begin{tabular}{l c c}
Species & $\rho_a$ (mol/l) & $N$ \\ \hline
C$_5$H$_{12}$ & 8.6048 & 332 \\ 
C$_6$H$_{14}$ & 7.5982 & 293 \\
C$_7$H$_{16}$ & 6.7823 & 261 \\
C$_8$H$_{18}$ & 6.1129 & 236 \\
C$_9$H$_{20}$ & 5.5677 & 215 \\
C$_{10}$H$_{22}$ & 5.1063 & 196 \\
C$_{12}$H$_{26}$ & 4.3780 & 169 \\ \hline
\end{tabular}
\end{table}

For simulations with the CGenFF forcefield, we use the NAMD \cite{namd} program. The starting simulation cells are energy minimized and then subjected to over 1~ns of equilibration under constant $NVT$ conditions. The temperature of 298.15~K is maintained using a Langevin thermostat. During this equilibration phase, the bond between the
hydrogen and the parent heavy atom was held fixed and the time step for integrating the equations of motion was 2~fs. (The final configuration from this phase was
subsequently used as a starting point for simulation using AMOEBA.) In the next phase, to allow a clear comparison with the simulations using AMOEBA, we 
remove the bond constraints on the hydrogens and equilibrate the system over 0.5~ns with 1~fs time step. Finally in the production phase that lasted 1~ns, we archive configurations every 100 fs to harvest 10000 configurations in all. The Lennard-Jones interactions are terminated at 14~{\AA} by smoothly switching to zero starting at 13~{\AA}. (Consistent with the forcefield model \cite{cgenff}, we do not include any additional long-range LJ corrections.) Long-range electrostatic interactions are treated using particle mesh Ewald summations with a real space cutoff of 13~{\AA}
and a grid size of 0.5~{\AA} for the reciprocal space sum. 

For simulations with the AMOEBA forcefield, we use the TINKER molecular modeling package \cite{tinker}.  We use the Andersen thermostat to maintain the temperature and use a time step of 1~fs for integrating the equations of motion. The van~der~Waals interactions are terminated at 13~{\AA} by smoothly switching to zero starting at 12~{\AA}. (Once again, long-range LJ corrections are not included.) Long-range electrostatic interactions are treated using Ewald summations with a real space cutoff of 13~{\AA}. Because of computational limitations the equilibration and production times varied among the various systems. Equilibration (production) times were as follows: C5, 62.3~ps (124.3~ps);  C6, 58.8~ps (113.0~ps); C7, 56~ps (116.5~ps); C8, 51.5~ps (118~ps); C9, 51.5~ps (124~ps); C10, 51.5~ps (93~ps); and C12, 51.5~ps (85.8~ps). During the production phase, configurations were saved every 250~fs for analysis.  

\subsection{Test particle calculations}
The test particle calculations are performed in two steps. First,  in each of the archived frames, we lay a cubic grid of points and accept the point as a viable insertion site if the
closest distance between the point and the nearest heavy atom is greater than 2.9 {\AA}. (We note that even by 3.0~{\AA}, the methane-water interactions become repulsive, indicating substantial overlap between the molecules \cite{asthagiri:jcp2008}.) This set of accepted points is saved. For calculations with the AMOEBA model, we separately calculate the energy of the water molecule $U_1$ and save it for subsequent use. For a given frame we first find the total energy of the system ($U_n$). Then we consider each of the viable insertion sites and write out new configuration files with both the alkane and inserted water coordinates. We then run an energy calculation using TINKER to obtain the energy of the complex $U_{n+1}$ and calculate $\Delta U$ (Eq.~\ref{eq:pdt}) for further analysis. 

A similar procedure was followed for configurations obtained using CGenFF.  However, for the test energy calculations, it proves more efficient to calculate $\Delta U$ using our in-house program. For long-range electrostatics we use the generalized reaction field \cite{Hummer:grf1992,Hummer:grf1994a,Hummer:grf1994b} procedure for computational economy. (Test calculations with Ewald summations showed that the binding energy values using GRF and Ewald are the same to better than two decimal accuracy.) 

For simulations with AMOEBA we attempt 2197 insertions per configuration for a total of between $0.7\times10^6$ (C12)  to $1.1\times 10^6$ (C5) trial points. 
Note that for the given sample size, the most positive free energy we can estimate is about 13~$k_{\rm B}T$($=\ln 1.1\times 10^6$), a number that is substantially larger than the largest positive value encountered in this work. (Hence we expect good statistical convergence.) 
Since we had many more frames with CGenFF, we only sample 343 points per configuration for a total of about $3\times 10^6$ trial points.  For statistical analysis, 
we treat $\langle e^{-\beta \Delta U} \rangle_0$ obtained per frame as a random variable. Using the Friedberg-Cameron algorithm \cite{allen:error,friedberg:1970} we then obtain the mean across
the entire sample (of frames) and the statistical inefficiency of the sampling. From this information, and using standard variance propagation rules, we calculate the
$\beta\mu^{\rm ex}_{w|a} = -\ln \langle  e^{-\beta \Delta U} \rangle_0$ and the associated standard error of the mean. 

\section{Results and Discussion}

\subsection{Methane-water pair potential}
\begin{figure}[h!]
\includegraphics[width=3.25in]{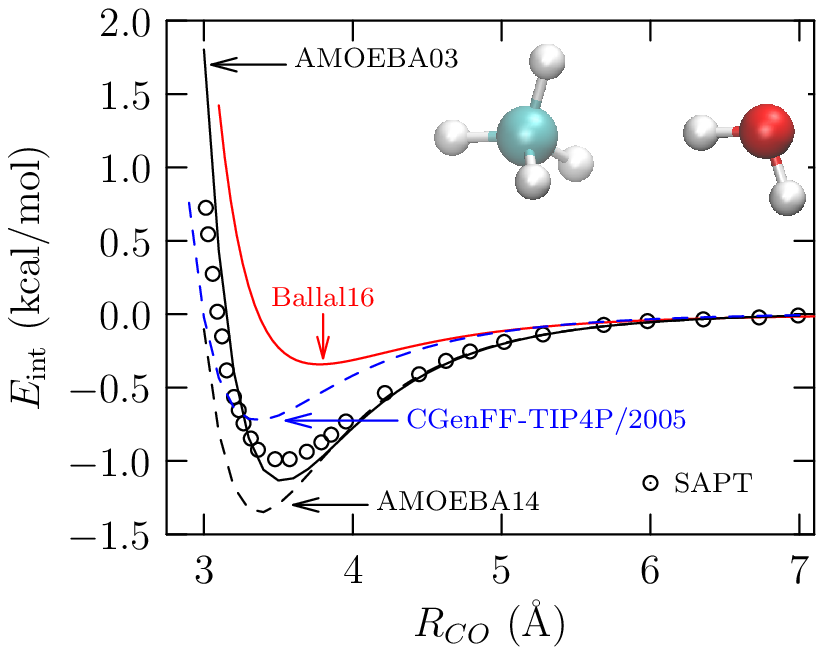}\\
\includegraphics[width=3.25in]{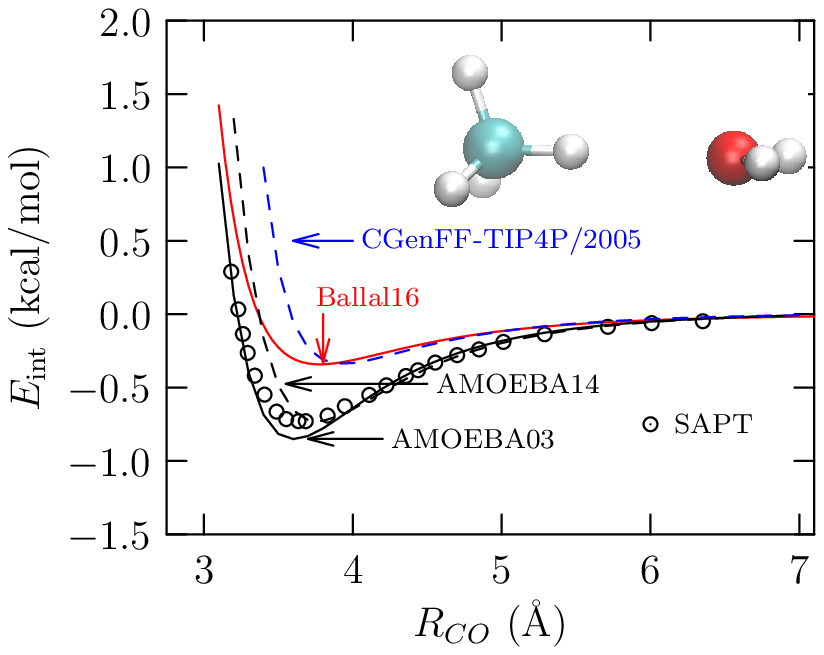}
\caption{Pair interaction profile between methane and water for the indicated configuration. \underline{Top panel}: water donates a proton to the electronegative carbon. \underline{Bottom panel}: oxygen accepts a proton from the methane. AMOEBA03: original AMOEBA water parameters \cite{amoeba03}; AMOEBA14: AMOEBA water parameters revised to better describe the liquid across a broad temperature range \cite{amoeba14}; CGenFF-TIP4P/2005: parameters based on standard Lorentz-Bertholet mixing using the alkane parameter set from CGenFF\cite{cgenff} and water based on TIP4P/2005\cite{tip4p}; Ballal16: parameters noted in Ref.~\onlinecite{ballal:jcp16}; methane-water interaction is based on a Lennard-Jones model with energy $\varepsilon_{MeW} = 172.05\, k_{\rm B}$ and collision diameter $\sigma_{MeW} = 3.365$~{\AA}.} 
\label{fg:fig1}
\end{figure}

Figure~\ref{fg:fig1} shows the methane-water pair interaction using different forcefields. The reference values are based on 
symmetry adapted perturbation theory (SAPT) results published in Ref.~\onlinecite{Szal:jcp05}. Calculations using Gaussian09 \cite{g09} at the 
MP2-level with aug-cc-pVTZ basis also lead to results that are close to the SAPT values. 
Notice that the AMOEBA03 model captures the location and magnitude of the minima (Fig.~\ref{fg:fig1}) quite well.  For AMOEBA14 \cite{amoeba14}, in the configuration 
with alkane donating the proton, the minimum is well captured, but the repulsive wall is shifted outwards; and in the configuration with alkane accepting a proton, the minimum is 
more pronounced. We note that the AMOEBA alkane parameters \cite{amoebaOrg} were also validated against AMOEBA03 water model, but the hydration free energy of methane in water is better predicted with the AMOEBA14 water model (Appendix~\ref{sc:methane}). 

The CGenFF-TIP4P/2005 model captures the anisotropy of the interaction, but relative to SAPT neither the location nor the magnitude of the potential minima are well described.  The pair-interaction profile based on Ballal et al.'s \cite{ballal:jcp16}  model 
is able to describe the experimental $\mu^{\rm ex}_{w|a}$ within a united atom framework, but once again neither the location nor the magnitude of the minima are captured. Overall, in comparison with the SAPT results, both the AMOEBA models are significantly better than either CGenFF-TIP4P/2005 or Ballal16 models. 


\begin{figure}[h!]
\includegraphics[width=3.25in]{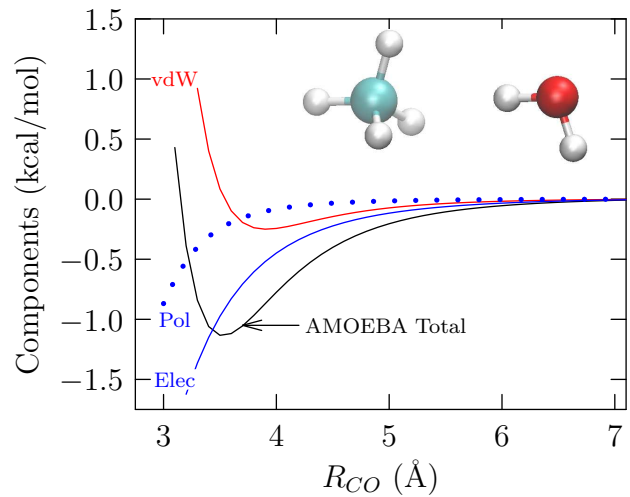}
\caption{The components of interaction profile according to the AMOEBA interaction model; water is described using the  AMOEBA03 paramaters. `vdw': van~der~Waals interaction; `pol': polarization (or induction) contribution; `elec': electrostatics based on the interacting distributed multipoles.}
\label{fg:fig2}
\end{figure}
Dissecting the interaction profile further (Fig.~\ref{fg:fig2}), we find that electrostatic interactions, and secondarily induction effects, account for much of the well-depth within the AMOEBA model. Although it is not prudent to infer a particular physical contribution as dominant on the basis of a forcefield description alone, with the background knowledge of the philosophy underlying the AMOEBA forcefield development,
the above results collectively suggest that electrostatic interactions and induction effects are important in alkane water interactions.

There is one paradoxical feature in the above results. For the interaction in the specific configurations (Fig.~\ref{fg:fig1}), the Ballal16 united atom model clearly does not 
describe the underlying physics correctly. Yet, this model captures $\beta\mu^{\rm ex}_{w|a}$ quite well. We address this issue next section in the context of the idea of coarse-graining. 

\subsection{Solvation of water in n-alkanes}\label{sc:solv}
\begin{figure}[h!]
\includegraphics[width=3.25in]{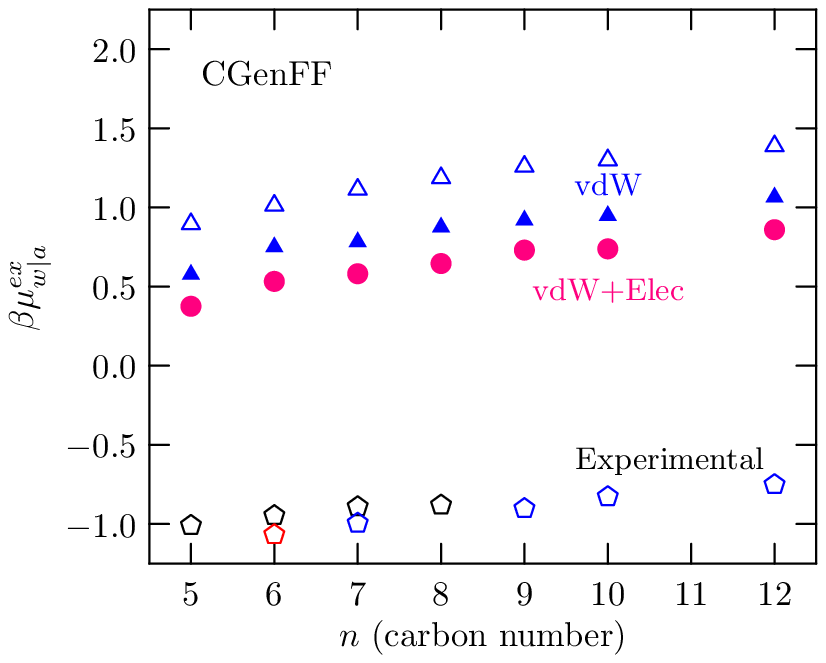} \\ 
\includegraphics[width=3.25in]{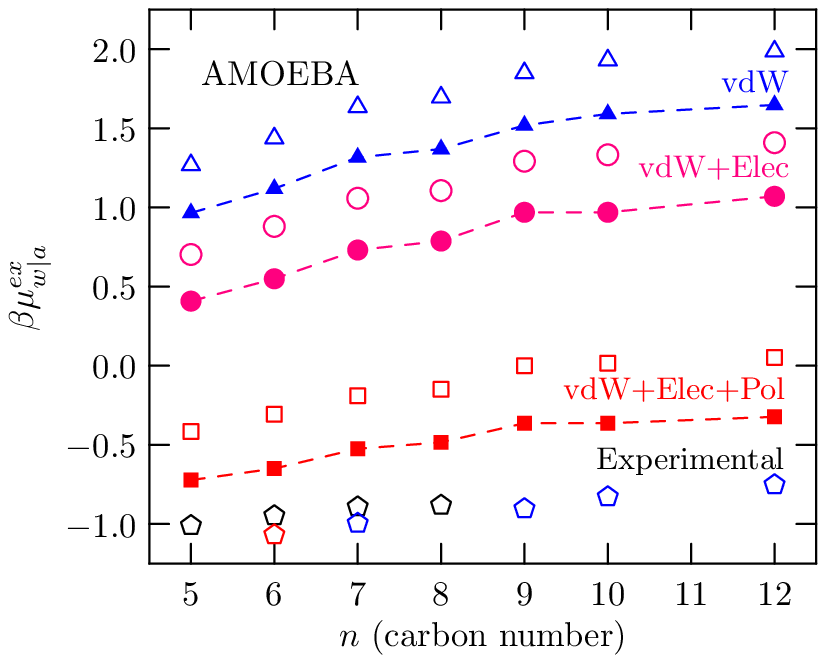}
\caption{The transfer free energy of water to n-alkanes C$_{n}$H$_{2n+2}$. \underline{Top panel}: Results using CGenFF. vdw:  only van~der~Waals interactions considered, 
and vdW+Elec, both van~der~Waals and electrostatics interactions included.  \underline{Bottom panel}: Results using AMOEBA03 (open symbols) and AMOEBA14 (corresponding closed symbols with dashed connecting lines to guide the eye) for water.  For AMOEBA, 
vdW+Elec+Pol (polarization) includes all the interactions.  Three sets of experimental data are shown using pentagons. Ref.\ \onlinecite{polak}, $n=5,6,7,8$; Ref.\ \onlinecite{roddy}, $n=6$; Ref.\ \onlinecite{schatzberg}, $n=7,9,10,12$. Standard error of the mean for $\beta\mu^{\rm ex}_{w|a}$ is comparable to the size of the symbol and hence is not shown.} \label{fg:dmu}
\end{figure}

Figure~\ref{fg:dmu} collects  $\mu^{\rm ex}_{w|a}$ for all the models considered in this work. 
Within CGenFF, $\mu^{\rm ex}_{w|a}$ is positive, contrary to experiments. However, the trend with respect to carbon number is correctly captured. Including electrostatic interactions modestly lowers the free energy, emphasizing the role of favorable water-alkane electrostatic interactions in the solvation process.

Both the AMOEBA water models lead to a negative $\mu^{\rm ex}_{w|a}$ and the trend with respect to carbon number is also correctly captured. The average deviation relative to experiments is about 0.25~$k_{\rm B}T$ for AMOEBA14 and 0.5~$k_{\rm B}T$ for AMOEBA03 water models. Within the AMOEBA model, electrostatic interactions
play a larger role than in CGenFF, lowering the free energy from the vdW-only value by about 0.5~$k_{\rm B}T$.  
Interestingly, whereas polarization played a secondary role in the case of a single methane-water pair (Fig.~\ref{fg:fig1}), \textit{multi-body polarization} effects in the condensed phase contribute substantially to lowering the free energy, with the net shift relative to the vdW+Elec case being about 1~$k_{\rm B}T$. 

\subsubsection{Coarse graining}

As noted above, Ballal et al.'s \cite{ballal:jcp16}  model is able to describe the experimental $\mu^{\rm ex}_{w|a}$ within a united atom framework. In this approach, the SPC/E parameters for water and the TraPPE parameter for alkanes are used, but the Bertholet rule for the effective cross-interaction energy is scaled by a factor 1.6 and the Lorentz rule for the effective collision diameter is scaled by a factor of 0.98. We note that earlier studies show that the results are quite insensitive to small changes in the collision diameter \cite{ballal:jcp14,emborsky:iec11,fouad:jceng14}.  (In a similar vein, with cross interaction parameters that depart from the Lorentz-Bertholet mixing rule, using the united atom 
TraPPE model for methane and TIP4P-2005 model for water, Ashbaugh and coworkers \cite{Ashbaugh:hh} were able to fit the
free energy of hydration of methane in water across a range of temperatures.)  We rationalize these results
by considering the methane-water pair interaction. 

\begin{figure}[h!]
\includegraphics[width=3.25in]{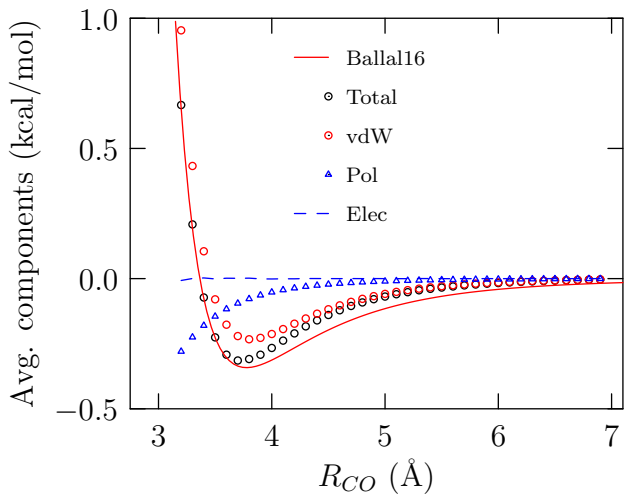}
\caption{Components of the unweighted angle-average interaction between methane and water using the AMOEBA03 model of water. (The AMOEBA14 model has a slightly lower minima that agrees better with Ballal16.) For each radial separation, on a grid we sample 30 points each for $\phi \in [0,2\pi]$ and for $cos(\theta)\in [-1,1]$.
For the orientation of water, with random numbers drawn from a uniform distribution, we choose 30 points from the set $\{\alpha, \cos(\beta), \gamma\}$. In all we sample 9000 configurations for each separation.} \label{fg:average}
\end{figure}

Figure~\ref{fg:average} shows the effective pair interaction between methane and water and its various components. 
This effective pair interaction is obtained by performing an \textit{unweighted} average over the relative configurations of a methane-water pair 
for a specified methane (C)-water (O) separation. (The location of the water relative to methane is specified by the spherical 
angles $\phi$ and $\theta$ and the rotation of the water relative to a fixed frame of reference is specified the Euler angles $\alpha$, $\beta$, and $\gamma$. 
As usual, we sample $\cos(\theta)$ and $\cos(\beta)$ to avoid singularities and ensure uniform sampling.) Remarkably, the effective methane-water interaction obtained
using AMOEBA is close to the Ballal16 united atom model. The effective interaction is dominated by van~der~Waals, and, to a lesser extent, polarization; 
 the effective electrostatic contribution is nearly zero for all separations. But this should not belie the importance of electrostatics, 
for without electrostatic interactions, we would not have induction effects (Fig.~\ref{fg:fig2}). 

\begin{figure}[h!]
\includegraphics[width=3.25in]{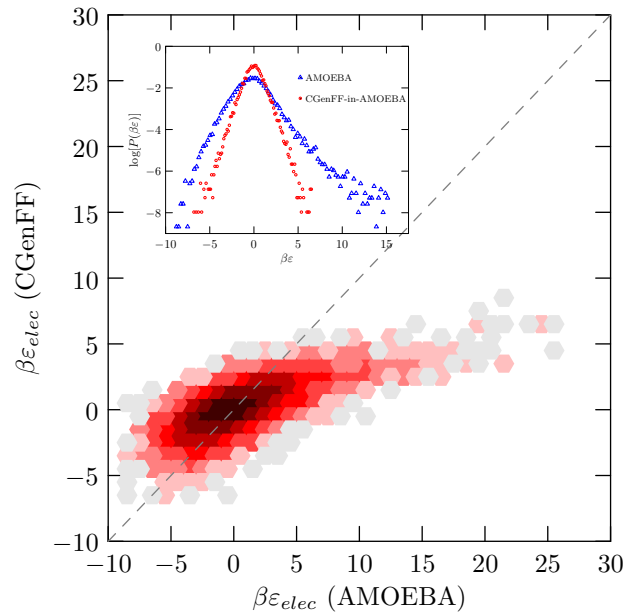}
\caption{Density plot of the joint distribution $\log P(\beta\varepsilon_{elec}[{\rm AMOEBA}], \beta\varepsilon_{elec}[{\rm CGenFF}])$ of electrostatic interaction energies obtained using AMOEBA and CGenFF.  The underlying configurations are the same for both. For CGenFF we use the SPC/E charges for water to preserve the mapping of the 3-site AMOEBA water model to a 3-site fixed charge water model. Every factor-of-$e$ change in the density relative to the mean corresponds to a change in shade. The inset is the projection of the distribution to the respective axis. For clarity, the inset x-axis is cut at $\beta\varepsilon = 15$.} \label{fg:elecAvg}
\end{figure}

The mapping of AMOEBA to the united atom model is an example of coarse graining. It is interesting to consider a similar mapping between  AMOEBA and CGenFF. 
To this end we focus on water solvation in C$_5$H$_{12}$. For configurations of n-pentane obtained using AMOEBA, we archive the distribution of electrostatic interaction
energies obtained from the test particle procedure. For the same relative orientation of the test water molecule, we also obtain the distribution of electrostatic interaction 
energy ($\beta\varepsilon_{elec}$) by replacing the AMOEBA distributed multipoles with CGenFF distributed monopoles (i.e.\ partial charges). For water, we choose the SPC/E partial charge distribution
to map the three-site AMOEBA model to another three-site water model. (Using TIP4P/2005 charges will not change the physical conclusion.) Fig.~\ref{fg:elecAvg} compares
the distribution of electrostatic contributions to the interaction energy. 
Please note that mean interaction energy and the states around the mean are comparably described on the basis of the AMOEBA distributed multipoles and the CGenFF-SPC/E distributed monopoles, but the energy distribution based on the latter model is sharper. Thus as we move away from the mean, the discrepancy in the description of the electrical interaction is higher, with AMOEBA-based electrostatic interaction energies displaying a greater variability.  The higher sensitivity of the distributed multipole description 
is, of course, anticipated. Importantly, in the lower energy wing, AMOEBA-based electrostatics predicts a more negative electrostatic contribution than CGenFF-SPC/E. 
It is for this reason, that the impact of the electrostatic effects using AMOEBA is higher than that using CGenFF (Fig.~\ref{fg:dmu}) . 

Returning to the problem of solvation, when we use the united-atom model for computing $\beta\mu^{\rm ex}_{w|a}$, effectively we are taking the average of Boltzmann-factors ($e^{-\beta\Delta U}$) with $\Delta U$ itself an angle-averaged interaction energy, whereas the correct approach would be to use equation Eq.~\ref{eq:pdt} with Boltmann-factors obtained first and then averaged over all configurations. The latter approach of course demands a complete description of the physics. The former approach may work, but 
it requires the effective interaction model be developed first, perhaps by fitting to available data. Our analysis makes it clear why it should not surprise us if such models lack transferrability. A similar effect undoubtedly underlies the mapping of a fully quantum description of interactions to AMOEBA or AMOEBA to CGenFF, but by preserving more of the physics, these latter coarse-graining steps are likely to allow better transferrability.  

With the caveat about forcefields and contributions noted above, our results (Fig.~\ref{fg:dmu}) show that electrostatic and induction effects play a decisive role in enhancing the solubility of water in alkanes, but these effects can be masked when we use united-atom models. The success of the AMOEBA model notwithstanding, relative to experiments an admittedly small discrepancy still persists. We consider some possible causes below.

\subsection{Polarization correction and liquid water-to-alkane transfer}

The TIP4P-2005 model has an artificially high dipole moment that is required for describing the liquid. Thus it is necessary to correct for this over-polarization in estimating the transfer free energy of TIP4P water from liquid to vapor. However, for transferring  the solute from vapor to the  apolar alkane phase, one need not apply any corrections and this is partly why we recast the experimental results in terms of vapor to liquid alkane transfer. However, we do make the assumption that the TIP4P-2005 model can be used to describe water-alkane interactions in a \textit{water-lean phase}. The present study shows that this is unlikely to be satisfactory. A thorough analysis of this issue is  left for future studies. We also note that no such polarization issues arise in the AMOEBA model. However, preliminary calculations using quasichemical theory \cite{asthagiri:pre03,weber:jcp10b} show that the liquid water to vapor transfer free energy using AMOEBA is more positive by about 0.3~kcal/mol than the experimental value for H$_2$O ($-6.32$~kcal/mol) at 298~K. Thus, if we compared liquid water to alkane transfer free energy, the results with AMOEBA will also be in substantial error. 

\subsection{Role of water dimerization}

Earlier \cite{ballal:jcp14} it has been suggested that clustering of water molecules in the alkane phase can potentially contribute to the partitioning of water in the alkane phase. Here we examine this suggestion. First note that as already discussed in Sec.~\ref{sc:expt} dimerization in the gas phase proved inconsequential to $\mu^{\rm ex}_{w|v}$. In the alkane phase, to lowest order in solute density the contribution of solute-solute pairing to the excess chemical potential is $2\rho_w^{(a)} \tilde{B}_2$  \cite{Vafaei:2014eb,Zhang:2014fr}, where $\tilde{B}_2$ is the osmotic second virial coefficient.  Quantum chemical calculations using B3LYP/6-311+G(d,p) and the SMD continuum solvation model \cite{smd} shows that  association is weaker in the solvent than in the gas-phase at 298~K. (The weakening of interaction in the alkane is analogous to the weakening of association between hydrophobes in water due to solute-water attractive interactions \cite{asthagiri:jcp2008,BenAmotz:jpcl15,Chaudhari:2015gm}.) 
On this basis we can infer that $|\tilde{B}_2| < |B_2|$ and thus dimerization ought to be inconsequential even in the alkane phase. 

\subsection{Role of quantum effects}

It is well known that mass effects are important in the phase behavior of water. For example,  at 298.15~K the free energy  of transferring D$_2$O from the saturated liquid to the saturated vapor is about 0.14~$k_{\rm B}T$ higher than that for H$_2$O. Capturing such mass effects at finite temperatures requires treating the motion of nuclei quantum mechanically. (Zero point calculations with the SMD model predict a value of 0.08~$k_{\rm B}T$, qualitatively consistent with the experimental results.) It is clear that the experimental transfer free energy values (Figs.~\ref{fg:dmu}) implicitly include the physics of these mass-dependent effects.  The AMOEBA03 or AMOEBA14 parameters must also implicitly account for these mass-dependent effects because some of the experimental condensed phase properties are included in the parameter development. However it is unclear whether such a description can apply for transfer across very dissimilar phases. In particular,  in transferring a water molecule from the (dilute) vapor phase to the alkane phase, we expect the rotational and vibrational modes of the water molecule to be  red-shifted because of interaction with the alkane. Consistent with this intuition, the zero-point energy of a water molecule in the alkane (modeled as a continuous dielectric medium) is lower by $0.1$~$k_{\rm B}T$ relative to that in the vapor. Including this correction over the classical statistical mechanical transfer free energy results (Figs.~\ref{fg:dmu}) improves the agreement with experiments. Our analysis of these mass effects is necessarily qualitative; a rigorous treatment of quantum effects associated with nuclear motion \cite{rossky:jcp85,guissani:jcp98} could prove rewarding, but such efforts are beyond the scope of this study. 

\section{Concluding Discussion}

In discussions of alkane-water interactions, dispersion interactions and, at times the contribution due to the polarizability of the alkane, are most often considered. 
However, for describing water solubility in alkanes with well benchmarked forcefields such as TraPPE, relying on dispersion interactions alone proves inadequate in capturing the experimentally observed water content in alkanes. The problem arises in modeling water-alkane cross interactions for which conventional mixing rules appear to fail.  Thus researchers have had to rely on empirical adjustments for the interactions in the alkane rich and alkane lean phases (for example, see Ref.\ \onlinecite{bolton:ms09}). These failures point to missing physics that needs to be accounted in modeling alkane-water interactions. 

Our calculations with the AMOEBA all-atom polarizable forcefield is able to describe the experimentally inferred transfer free energy of water from its vapor to a liquid alkane. 
But even with AMOEBA, relying on van~der~Waals interactions alone proves inadequate; one must include both electrostatics and induction effects to predict correctly the water content in alkanes.  Since the parameters in the AMOEBA forcefield rely extensively on first principles calculations, especially in deriving the atomic polarizability and  the distributed multipoles, we conclude that electrostatic interactions and induction effects are important physical ingredients in modeling alkane-water interactions. 

We establish the connection between the AMOEBA model and the united atom model  with pair-interaction that is empirically adjusted to capture water solubility in alkanes.  
We show that the unweighted configurationally averaged interaction between methane and water using AMOEBA is in close agreement with the united atom model.  
The average interaction is dominated by van~der~Waals and induction effects, the sum of which can be fit to a revised van~der~Waals potential. But 
the approximately zero electrostatic contribution should not belie the importance of electrostatics in this interaction. Thus since the united atom models are effective interactions,
it is not surprising if we have to re-optimize the interaction between different chemical species in different states of aggregation. Similar comments apply in mapping 
the physical electrostatic interactions to one described by partial charges alone.  Thus using fixed-charge water models such as TIP4P/2005 in a water-lean phase, such as in alkanes or in the interior of biological molecules, may also incur errors, especially if one is interested in the thermodynamics of transfer of the water molecule into such phases. More thoroughly exploring these issues are left for future studies.

\section{Acknowledgements}
DA, AVP, WGC are grateful for financial support from the Robert A.\ Welch Foundation (Grant No. C-1241),  and the Rice University Consortium for Processes in Porous Media (Houston, TX, USA).  DB was supported by the Division of Material Sciences and Engineering, Office of Basic Energy Sciences, U.S. Department of Energy, under Contract No. W-7405-430 ENG-82 with Iowa State University. We thank B.\ Xue, D.\ B.\ Harwood, and J.\ I.\ Siepmann for their comments 
on Ref.~\onlinecite{ballal:jcp14} which led us to correct that work \cite{ballal:jcp16} and focus on the vapor-to-liquid transfer process. 
This research used resources of the National Energy Research Scientific Computing Center, a DOE Office of Science User Facility supported by the Office of Science of the U.\ S.\ Department of Energy under Contract No.\ DE-AC02-05CH11231.

\section{Appendix}
\subsection{Hydration of methane within AMOEBA}\label{sc:methane}
To complement the study of water solvation in alkane, we also tested the hydration of methane. To this end, we simulated a box of 512 water molecules and obtained the hydration free energy of methane using Eq.~\ref{eq:pdt}.  The solvent was simulated at a temperature of 298.15~K and the experimental density of 0.997 gms/cc. 
The temperature was maintained using an Andersen thermostat. The dispersion interactions were smoothly switched to zero between 9~{\AA} and 10~{\AA}. Electrostatic interactions were treated using Ewald summations with the real-space interaction cut at 10~{\AA}. The time step for integrating the equations of motion was 1~fs.

For either AMOEBA03 or AMOEBA14, we took a pre-equilibrated box of water molecules and equilibrated it further for 250~ps. The subsequent production phase lasted 
an additional 625~ps.  In the production phase configurations were saved every 250~fs to harvest a total of 2500 configurations. The particle insertion calculation was
conducted as described in the main text. For methane hydration, for each viable insertion site, we consider five (5) random orientations of the methane molecule. 
For simplicity we use the single equilibrium geometry of methane, although in the original  validation study \cite{amoebaOrg} the alkane was flexible. 

Table~\ref{sc:tableMe} collects the results of our studies on methane hydration. We find that the hydration free energy is well estimated using the AMOEBA14 water model, while the AMOEBA03 model underpredicts the hydration free energy by 20\%.  We must note that in Ref.~\onlinecite{amoebaOrg}, the hydration free energy of methane obtained with the AMOEBA03 water model was in near perfect agreement with experiments. The reasons for the discrepancy between our result and the published value is not clear. 
Perhaps the differences in system size 216 versus 512 water molecules and the assumption of rigid structure of the alkane may explain some of the deviation.
\begin{table}[h!]
\caption{Simulated properties of neat water and the hydration free energy of methane on the basis of including only van~der~Waals ($\mu^{\rm ex}_{v}$), van~der~Waals plus electrostatics ($\mu^{\rm ex}_{ve}$), and van~der~Waals, electrostatics, and polarization contributions ($\mu^{\rm ex}_{vep}$). The free energies are in kcal/mol. The experimental hydration free energy is from Ref.\ \onlinecite{wilhelm:cr77}.}\label{sc:tableMe} 
\begin{tabular}{c c c c}
Property  & AMOEBA03 &AMOEBA14 & Expt. \\ \hline
$\mu^{\rm ex}_{v}$ & $2.0\pm0.03$ & $2.5\pm0.05$ & --- \\
$\mu^{\rm ex}_{ve}$ & $1.9\pm0.03$ & $2.3\pm0.05$ & --- \\
$\mu^{\rm ex}_{vep}$ & $1.6\pm0.03$ & $2.0\pm0.05$ & 2.00 \\ \hline
\end{tabular}
\end{table}
The results in Table~\ref{sc:tableMe} suggest that electrostatics and polarization also play an important role in the hydration of methane; we expect a similar behavior for other alkanes as well.

\newpage


\end{document}